\documentclass[12pt]{article}
\newcounter{saveeqn}
\usepackage{epsfig}

\newcommand{\alpheqn}{\addtocounter{equation}{1}\setcounter{saveeqn}%
{\value{equation}}\setcounter{equation}{0}%
\renewcommand{\theequation}{\mbox{\arabic{saveeqn}\alph{equation}}}}
\newcommand{\reseteqn}{\setcounter{equation}{\value{saveeqn}}%
\renewcommand{\theequation}{\arabic{equation}}}

\renewcommand{\vec}[1]{\mbox{\boldmath $ #1$}}
\newcommand{\captionfonts}{\small \sl}
\makeatletter  
\long\def\@makecaption#1#2{%
  \vskip\abovecaptionskip
  \sbox\@tempboxa{{{\bf #1:} \captionfonts #2}}%
  \ifdim \wd\@tempboxa >\hsize
    { {\bf#1:} \captionfonts#2\par}
  \else
    \hbox to\hsize{\hfil\box\@tempboxa\hfil}%
  \fi
  \vskip\belowcaptionskip}
\makeatother   

\title{\bf Patterns of Convection in Rotating Spherical Shells}
\author{by \sf R. Simitev and F.H. Busse }
\date{\small Institute of Physics, University of Bayreuth, D-95440 Bayreuth, Germany}

\begin{document}
\maketitle

\begin{abstract}
New J. Phys., 5, pp. 97.1-97.20, DOI:10.1088/1367-2630/5/1/397, 2003. \\

Patterns of convection in internally heated, self gravitating rotating
spherical fluid shells  are investigated through numerical
simulations. While turbulent
states are of primary interest in planetary and stellar applications
the present paper emphasizes more 
regular dynamical features at Rayleigh numbers not far above
threshold which are similar to those which might be observed in
laboratory or space experiments. Amplitude vacillations and spatial modulations of
convection columns are  common features at moderate and large Prandtl
numbers. In the low Prandtl number regime equatorially attached
convection evolves differently with increasing Rayleigh number and
exhibits an early transition into a chaotic state. Relationships of
the dynamical features to coherent structures in fully turbulent
convection states are emphasized.     
\end{abstract}

\section{Introduction}

Thermal convection in a rotating spherical fluid shell represents one
of the fundamental systems on which a considerable part of the
understanding of many observed geophysical, planetary and astrophysical
phenomena is based. It is generally believed that the magnetic field
exhibited by celestial bodies is generated in their interiors by
convection driven by thermal and compositional buoyancy. Cloud
patterns and the differential rotation seen at the surface of the
major planets are likely to be connected with convection in the deep
atmospheres of those planets (Busse, 1976, 1994). Considerable
attention has therefore been devoted to the  study of dynamical
properties of rotating spherical shells. For recent reviews we refer
to the papers by Zhang and Busse (1998) and Busse (2002).

\begin{figure}[ht]
\begin{center}
\epsfig{file=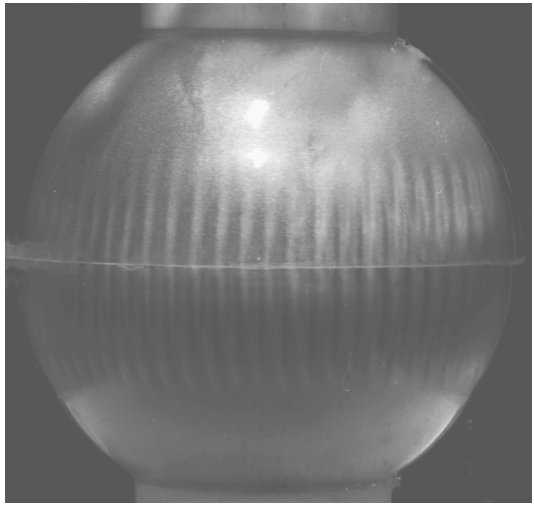,width=7cm}
\end{center}
\caption{Banana cells in a thin rotating spherical fluid shell cooled
from within. The motion is made visible by a suspension of flakes which
become aligned with the shear.}
\label{f1}
\end{figure}

Because a radially directed buoyancy force can not be easily realized
in Earth-bound laboratories, experimenters had to resort to the use of
centrifugal accelerations (Busse and Carrigan, 1976) or use
spacecrafts for their experiments (Hart {\it et al.} 1986). Fortunately, it
has turned out that the component of gravity perpendicular to the axis
of rotation is much more important than the component parallel to
it. The centrifugal force is thus a convenient substitute since
through the reversal of the applied temperature gradient the same
buoyancy can be obtained as for the inward directed component of
gravity perpendicular to the axis. Moreover, the centrifugal force
increases with the distance from the axis just as the case of the
gravity field of a spherical body of constant density. The use of
electric fields (Hart {\it et al.} 1986) or magnetic fields with ferrofluids 
(Rosenzweig {\it et al.} 1999) leads to rather
different radial dependences of the effective gravity. It must be
admitted, however, that the absence of an axial component of gravity in
experiments employing the centrifugal force leads to a thermal wind
type flow in the basic state. But this can be minimized when the
Coriolis parameter is sufficiently large or when a thin spherical
shell is used as for example in the experimental realization of
convection shown in figure \ref{f1}.

In this paper results from a numerical study will be reported that
have been obtained with essentially the same computer code as employed
in earlier studies focusing on turbulent convection (Tilgner and
Busse, 1997) and on the generation of magnetic fields (Grote and
Busse, 2001). The mathematical formulation and the description of the
numerical method given in section 2 can thus be kept brief. In section
3 a short introduction to the results of linear theory for the onset
of convection will be given. In section 4 properties of columnar
convection at Prandtl numbers of the order unity will be described and
some animations of the time dependence will be presented. Equatorially
attached convection flows are the subject of section 5. Their
properties at finite amplitudes are studied here for the first
time. Because of the required high numerical resolution in time as
well as in space the low Prandtl number regime of fully nonlinear
convection has posed a significant computational challenge until now.
 As will be shown the 
continuity of the convection heat flux requires that equatorially
attached convection must be joined by columnar convection as the
Rayleigh number increases significantly above its critical value. A
concluding discussion will be given in section 6.

\section{ Mathematical Formulation and Numerical \\ \hspace*{0.05cm} Method}

For the description of finite amplitude convection in rotating
spherical fluid shells we follow the standard formulation  used  by
many authors Zhang and Busse(1987), Ardes {\it et al.} (1997),
Zhang(1991, 1992). It is assumed that a general static state exists
with the temperature distribution $T_S = T_0 - \beta d^2 r^2 /2 +
\Delta T \eta r^{-1} (1-\eta)^{-2}$ where $\eta$ denotes the ratio of
inner to outer radius of the shell and $d$ is its thickness. $\Delta
T$ is the temperature difference between the boundaries in the special
case $\beta =0$ of vanishing internal heat sources. The gravity field
is given by $\vec g = - \gamma d \vec r$ where $\vec r$ is the
position vector with respect to the center of the sphere and $r$ is
its length measured in units of $d$. In addition to the length $d$,
the time $d^2 / \nu$ and  the temperature $\nu^2 / \gamma \alpha d^4$
are used as scales for the dimensionless description of the problem
where $\nu$ denotes the kinematic viscosity of the fluid and $\kappa$
its thermal diffusivity. The density is assumed to be constant except
in the gravity term where its temperature dependence given by $\alpha
\equiv ( d \varrho/dT)/\varrho =$ const. is taken into account.
The basic equations of motion and the heat equation for the deviation
$\Theta$ from the static temperature distribution are thus given by
\alpheqn
\begin{equation}
\label{1a}
\partial_t \vec{u} + \vec u \cdot \nabla \vec u + \tau \vec k \times
\vec u = - \nabla \pi +\Theta \vec r + \nabla^2 \vec u
\end{equation}
\begin{equation}
\label{1b}
\nabla \cdot \vec u = 0
\end{equation}
\begin{equation}
\label{1c}
P(\partial_t \Theta + \vec u \cdot \nabla \Theta) = (R_i+R_e\eta
r^{-3}(1-\eta)^{-2}) \vec r \cdot \vec u + \nabla^2 \Theta 
\end{equation}
\reseteqn
where the Rayleigh numbers $R_i$ and $R_e$, the Coriolis parameter $\tau$
and  the Prandtl number $P$ are defined by  
\reseteqn
\begin{displaymath}
\label{1d}
R_i = \frac{\alpha \gamma \beta d^6}{\nu \kappa} ,  \enspace R_e =
\frac{\alpha \gamma \Delta T d^4}{\nu \kappa} , \enspace \tau =
\frac{2 \Omega d^2}{\nu} , \enspace P = \frac{\nu}{\kappa}. 
\end{displaymath}
 Since
the velocity field $\vec u$ is  solenoidal the general representation
in terms of poloidal and toroidal components can be used, 
\begin{displaymath}
\label{1e}
\vec u = \nabla \times ( \nabla v \times \vec r) + \nabla w \times
\vec r .
\end{displaymath}
By multiplying the (curl)$^2$ and the curl of equation (\ref{1a}) by
$\vec r$ we obtain two equations for $v$ and $w$,  
\alpheqn
\begin{equation}
\label{2a}
[( \nabla^2 - \partial_t) L_2 + \tau \partial_{\phi} ] \nabla^2 v +
\tau Q w - L_2 \Theta  = - \vec r \cdot \nabla \times [ \nabla \times
( \vec u \cdot \nabla \vec u )] 
\end{equation}
\begin{equation}
\label{2b}
[( \nabla^2 - \partial_t) L_2 + \tau \partial_{\phi} ] w - \tau Qv
= \vec r \cdot \nabla \times ( \vec u \cdot \nabla \vec u) 
\end{equation}
\reseteqn
where $\partial_t$ and $\partial_{\phi}$ denote the partial
derivatives with respect to time $t$ and with respect to the  angle
$\phi$ of a spherical system of coordinates $r, \theta, \phi$
and where the operators $L_2$ and $Q$ are defined by  
\alpheqn
\begin{equation}
\label{3a}
L_2 \equiv - r^2 \nabla^2 + \partial_r ( r^2 \partial_r)
\end{equation}
\begin{equation}
\label{3b}
Q \equiv r \cos \theta \nabla^2 - (L_2 + r \partial_r ) ( \cos \theta
\partial_r - r^{-1} \sin \theta \partial_{\theta}). 
\end{equation}
\reseteqn
Stress-free boundaries with fixed temperatures are most often assumed,
\begin{equation}
\label{4}
v = \partial^2_{rr}v = \partial_r (w/r) = \Theta = 0 
\end{equation}
\begin{displaymath}
\mbox{ at } \enspace r=r_i \equiv \eta / (1- \eta) \enspace \mbox{ and at }
\enspace r=r_o = (1-\eta)^{-1} 
\end{displaymath}
but sometimes no-slip boundaries are used in which case the conditions
$\partial^2_{rr} v = \partial_r (w/r)=0$ must be replaced by
$\partial_r (rv) = w = 0$.  For
simplicity we shall assume the previously used value $\eta = 0.4$ unless
indicated otherwise. The 
numerical integration of equations (\ref{1c}) and (2) together with boundary
conditions (\ref{4}) proceeds with the pseudo-spectral method  which is
based on an expansion of all dependent variables in spherical
harmonics for the $\theta , \phi$-dependences, i.e.  
\begin{equation}
\label{5}
v = \sum \limits_{l,m} V_l^m (r,t) P_l^m ( \cos \theta ) \exp \{ im
\phi \} 
\end{equation}
and analogous expressions for the other variables, $w$ and
$\Theta$. $P_l^m$ denotes the associated Legendre functions. For the
$r$-dependence expansions in Chebychev polynomials are used. Spherical
harmonics up to the order of $l=96$ and up to 41 collocation points in
radial direction have been used. A time step as low as $10^{-6}$ was
required for the computations of the small Prandtl number cases
reported in section 5. For further details see Busse {\it et al.}
(1998), Tilgner and Busse (1997). 

Benchmark comparisons for stationary solutions of equations (\ref{1c})
and (2)
together with conditions  (\ref{4}) when the dynamo process is included have
been done in the past (Christensen {\it et al.}, 2001).  

\section{Onset of Convection in Rotating Spherical Shells}

\begin{figure}[t]
\begin{center}
\hspace*{1cm}\epsfig{file=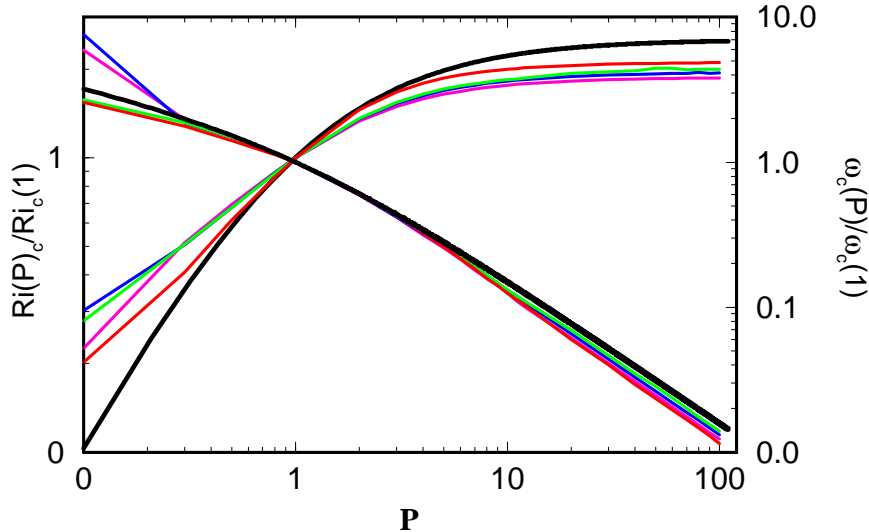,width=7cm,angle=-90}
\end{center}
\caption{Critical Rayleigh number $R_{ic}$ and frequency $\omega_c$
  (right ordinate) as a function of the Prandtl number $P$
in the case $\eta = 0.4$ for the Coriolis numbers $\tau = 5 \cdot
10^3$ (pink), $10^4$ (blue), $1.5 \cdot 10^4$ (green) and
$10^5$ (red). The thick black line corresponds to expressions
(\ref{6a}) and (\ref{6c}) .
The reference value $R_{ic} (1)$ equals $80296$, $190140$, $318395$, $3665919$ for
$\tau = 5 \cdot 10^3, 10^4, 1.5 \cdot 10^4, 10^5$, respectively.
For expression (\ref{6a}) $R_{ic} (1)$ is given by $0.67543 \cdot \tau^{4/3}$.}
\label{f2}
\end{figure}

A rough idea of the dependence of the critical Rayleigh number $R_{ic}$ on
the parameters of the problem in the case of internal heating can be gained
from the expressions derived from the annulus model (Busse, 1970)
\alpheqn
\begin{equation}
\label{6a}
R_{ic} = 3 \left( \frac{P \tau }{1+P} \right)^{\frac{4}{3}} ( \tan
\theta_m)^{\frac{8}{3}} r_m^{-\frac{1}{3}} 2^{-\frac{2}{3}}   
\end{equation}
\begin{equation}
\label{6b}
m_c = \left( \frac{P \tau}{1+P} \right)^{\frac{1}{3}} ( r_m \tan
\theta_m )^{\frac{2}{3}} 2^{-\frac{1}{6}}  
\end{equation}
\begin{equation}
\label{6c}
\omega_c = \left( \frac{\tau^2}{(1+P)^2P} \right)^{\frac{1}{3}}
2^{-\frac{5}{6}} 
(\tan^2 \theta_m / r_m )^{\frac{2}{3}}
\end{equation}
\reseteqn
where $r_m$ refers to the mean radius of the fluid shell, $r_m = (r_i + r_o)/2$,
and $\theta_m$ to the corresponding colatitude, $\theta_m =$ arcsin$(r_m(1-\eta))$.
The azimuthal wavenumber of the preferred mode is denoted by $m_c$ and the
corresponding angular velocity of the drift of the convection columns in the
prograde direction is given by $\omega_c / m_c$.
In figure \ref{f2} the expressions (\ref{6a}, \ref{6c}) are compared with accurate
numerical values which indicate that the general  trend is well
represented by expressions (\ref{6a}, \ref{6c}). 
The same property holds for $m_c$. For a rigorous asymptotic analysis
including the radial dependence we 
refer to Jones {\it et al.} (2000). In the following we shall continue to
restrict the attention to the case $R_e=0$ and thus we shall drop the
subscript $i$ of $R_i$.

There is a second mode of convection which becomes preferred at onset
for sufficiently low Prandtl numbers. It is characterized by convection
cells attached to the equatorial part of the outer boundary not unlike
the ``banana cells'' seen in the narrow gap experiment of figure
\ref{f1}. The equatorially attached mode actually represents an inertial wave
modified by the effects of viscous dissipation and thermal
buoyancy. An analytical description of this type of convection can
thus be attained through the introduction of viscous friction and
buoyancy as perturbations as has been done by Zhang (1994) for the case
of stress-free as well as for no-slip boundaries (Zhang,
1995). According to Ardes {\it et al.} (1997) equatorially attached
convection is preferred at onset for $\tau < \tau_l$ where $\tau_l$
increases in proportion to $P^{1/2}$. 

In the derivation of the
asymptotic relationships (6) the exact nature of the boundary
condition for the tangential component of the velocity vector at the
spherical surface does not enter. In the limit of infinite $\tau$ the
conditions of onset of convections are the same for no-slip and for
stress-free conditions. In the following we shall restrict the attention,
however, to the case of stress-free boundaries in order to avoid the
complications arising from Ekman boundary layers at no-slip surfaces.

\section{Evolution of Convection Columns at Moderate Prandtl Numbers}

\begin{figure}[h]
\begin{center}
\epsfig{file=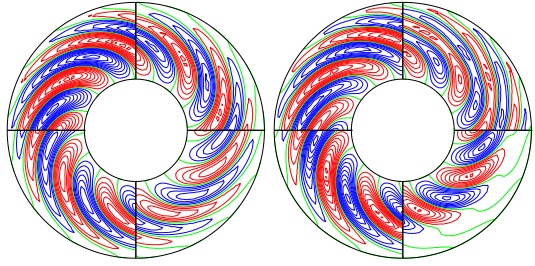,width=7cm}
\end{center}
\caption{Lines of constant $r \frac{\partial v}{\partial \phi}$ in the
  equatorial plane in the case $\tau=3 \cdot 10^4$, $P=0.1$ for $R=3
  \cdot 10^5$ (left) and for $R=3.5 \cdot 10^5$ (right). The four
  sections are a quarter of the vacillation period, $t_p=0.0864$
  (left) and $t_p=0.0124$ (right), apart with time progressing in the
  clockwise sense. } 
\label{f3}
\end{figure}
In general the onset of convection in rotating fluid spheres occurs
supercritically. As long as the convection assumes the form of shape
preserving travelling thermal Rossby waves as described by linear theory,
its azimuthally averaged properties are time independent. In fact, as seen
from a frame of reference drifting together with the convection columns
the entire pattern is steady. A differential rotation is generated through
the action of the Reynolds stress. The latter is caused by the spiralling
cross section of the columns which persists as a dominant feature at
moderate Prandtl numbers far into the turbulent regime. The plots of the
streamlines $r \frac{\partial v}{\partial \phi}= \mbox{const.}$ in the
equatorial plane shown in figure \ref{f3} give 
a good impression of the spiralling nature of the columns. 

A true time dependence of convection develops in the form of vacillations
after a subsequent bifurcation. First the transition to amplitude
vacillations occurs in which case just the amplitude of convection varies
periodically in time as exhibited in the left plot of figure \ref{f3}. At a
somewhat higher Rayleigh number shape vacillations become noticeable
which are characterized by periodic changes in the structure of the
columns as shown in the right plot of figure \ref{f3}. The outer part of the
columns is stretched out, breaks off and decays. The thinning and
subsequent breakup can actually be best seen in movie 1. The tendency
towards breakup is caused by the fact that the local frequency of
propagation varies with distance from the axis according to expression
(6c) after $\theta_m$ has been replaced by the local colatitude $\theta$.

\begin{figure}[t]
\begin{center}
\epsfig{file=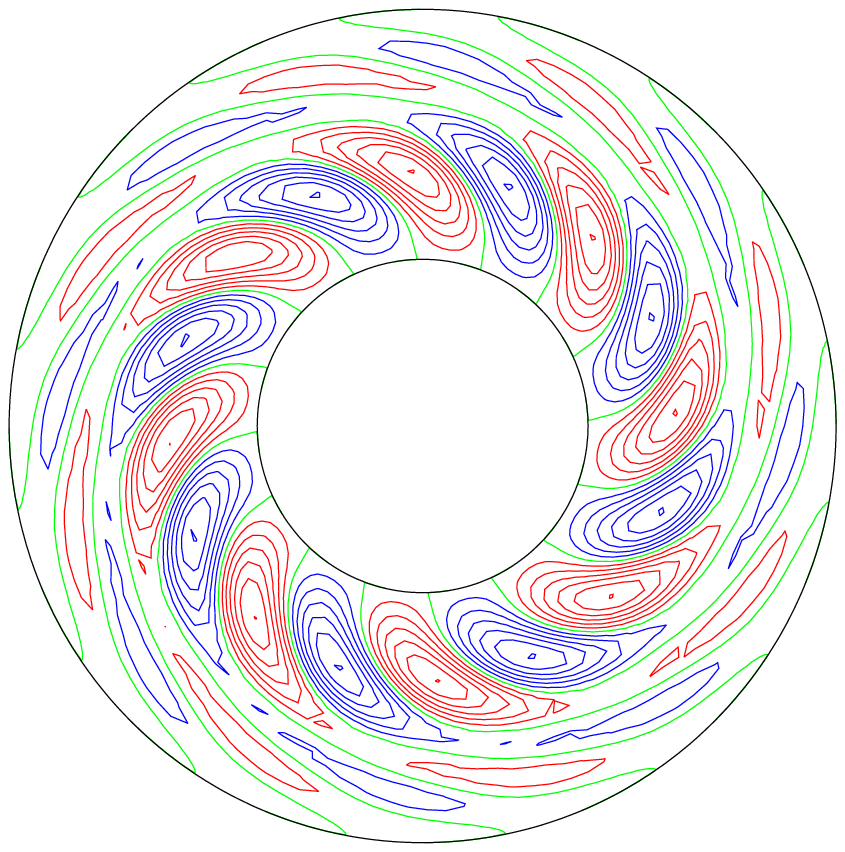,width=3.5cm} \hspace{0.5cm}
\epsfig{file=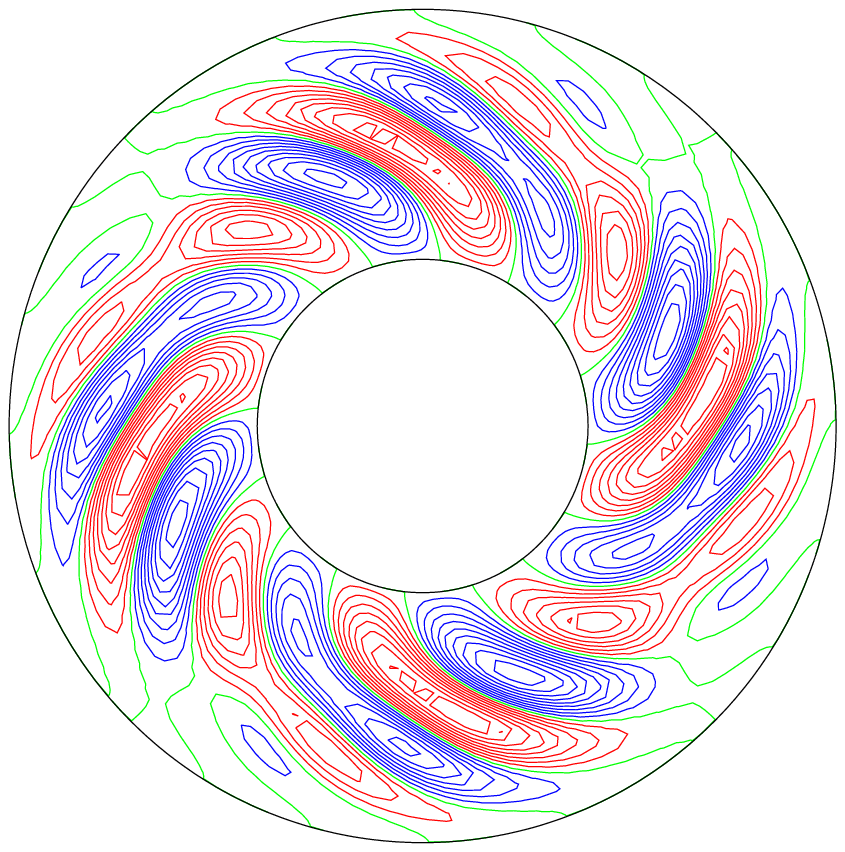,width=3.5cm} \hspace{0.5cm}
\epsfig{file=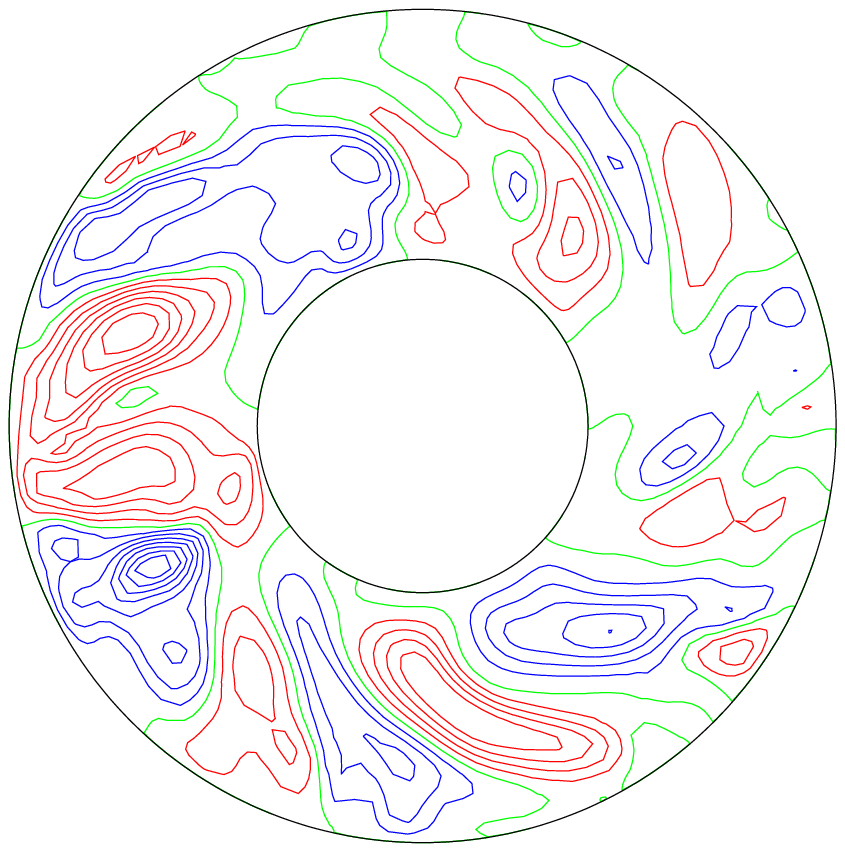,width=3.5cm} \\[5mm]
Movies 1, 2 and 3.
\end{center}%\caption{Movie 1}
\end{figure}
\begin{figure}[t]
\begin{center}
\hspace*{2.5cm}\epsfig{file=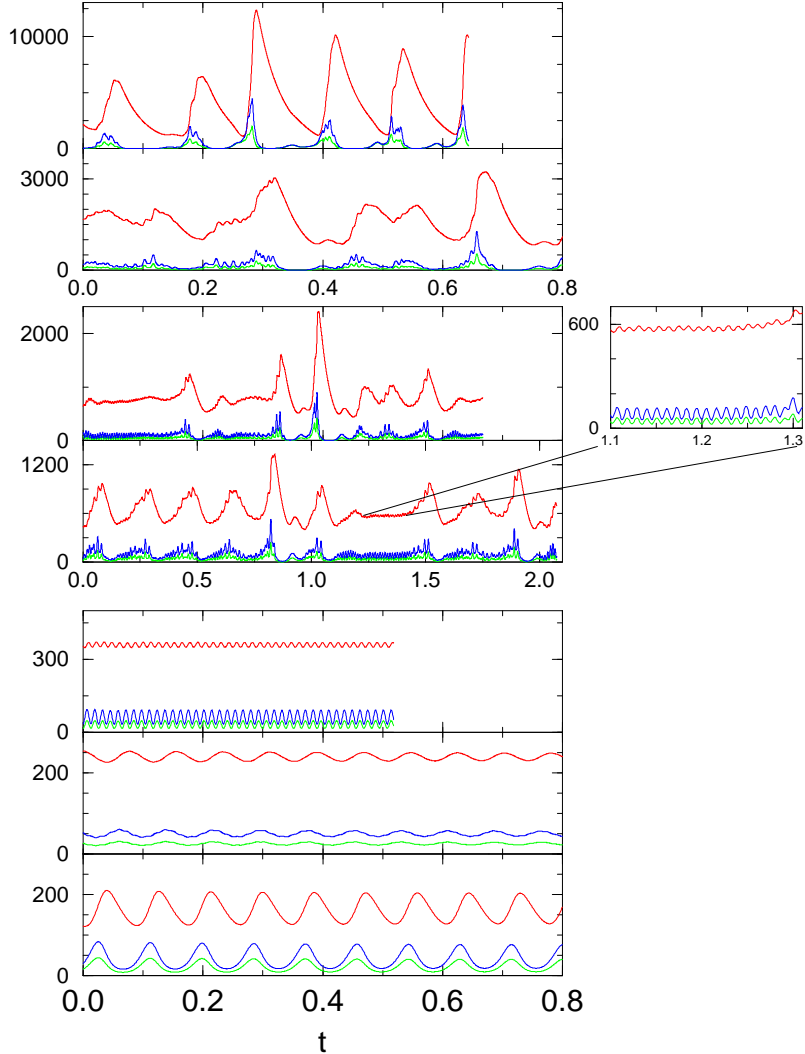,width=14cm,angle=-90}
\end{center}
\caption{Time series of energy densities of convection in the case
$\tau=3 \cdot 10^4$, $P=0.1$ for $R=3 \cdot 10^5$, $3.3 \cdot 10^5$,
$3.5 \cdot 10^5$, $3.8 \cdot 10^5$, $4 \cdot 10^5$, $4.5 \cdot 10^5$,
$5 \cdot 10^5$ (from bottom to top). Red, blue and green lines indicate
$E_t^m$, $E_t^f$, and $E_p^f$ respectively. The critical Rayleigh
number for onset of convection is $R_c=222518.$ $E_p^m$ is smaller by
more than an order of magnitude than the other energy densities and
has not been plotted for this reason.}
\label{f4}
\end{figure}

The two types of vacillations also differ significantly in their
frequencies of oscillation. This is evident from the time records of the
energy densities of convection which have been plotted in figure \ref{f4}.
The various components of the energy densities are defined by
\begin{equation}
\label{7}
E_p^m = \frac{1}{2} \langle \mid \nabla \times ( \nabla \bar v \times \vec r 
) \mid^2 \rangle , \quad E_t^m = \frac{1}{2} \langle \mid \nabla \bar w \times
\vec r \mid^2 \rangle
\end{equation}
\begin{equation}
\label{8}
E_p^f = \frac{1}{2} \langle \mid \nabla \times ( \nabla \check v \times \vec r) 
\mid^2 \rangle , \quad E_t^f = \frac{1}{2} \langle \mid \nabla \check w \times
\vec r \mid^2 \rangle
\end{equation}
where $\bar v$ refers to the azimuthally averaged
component of $v$ and $\check v$ is given by $\check v = v - \bar v $. 

\begin{figure}[t]
\begin{center}
\epsfig{file=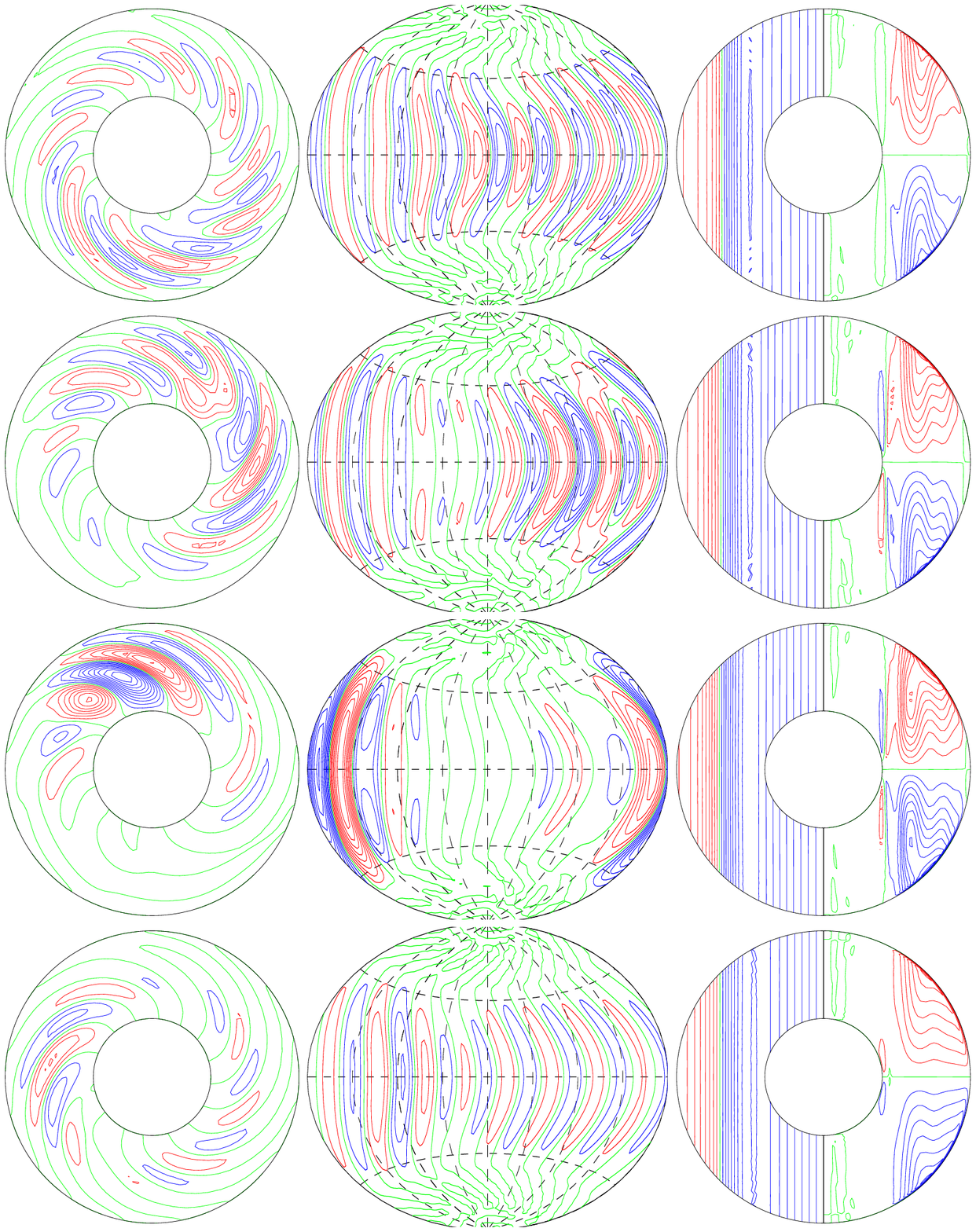,width=11.5cm}
\end{center}
\caption{Sequence of plots equidistant in time (from top to bottom with
 $\Delta t=0.05$) for $R=3.8\cdot 10^5$, $\tau=3 \cdot 10^4$,
 $P=0.1$. The left column shows streamlines, $r \frac{\partial v}{
 \partial \phi}=const.$, in the 
equatorial plane and in the middle column lines of constant radial
velocity $u_r$ on the mid surface, $r=r_i+0.5$, are shown. The
left halves of the circles of the right column show lines of constant
$u_\phi$ which is  the azimuthally averaged azimuthal component of the
velocity field. The right halves show streamlines of the axisymmetric
meridional circulation.}
\end{figure}
As the Rayleigh number is increased further a fairly sudden transition
into a chaotic regime occurs where convection has become strongly
inhomogeneous in space and in time. A typical sequence of plots is shown
in figure 5 which covers about one period of the the relaxation cycles
seen in the time record for $R=3.8 \cdot 10^5$ in figure \ref{f4}. In
contrast to the 
more common relaxation oscillations encountered at higher Rayleigh
numbers, see, for example, the time record for $R=5 \cdot 10^5$ in figure \ref{f4},
convection does not die off entirely at any time during the cycle. But the
interaction between convection and differential rotation appears to be
similar. As the amplitude of convection as measured by $E_p$ and $E_t$
grows the differential rotation generated by the Reynolds stress grows as
well with just a small delay in time. When the differential rotation
reaches a critical level the convection columns become disrupted and their
amplitude decays. Subsequently the differential rotation decays as well on
the time scale of viscous diffusion. It is typical for this type of
relaxation cycles that the viscous decay is shorter than the growth time
of the differential rotation in contrast to the relaxation oscillations at
higher values of $R$.
The regime of relaxation oscillations is interrupted once in a while by a
return to the more regular regime of shape vacillations as shown in the
inserted enlargement of the time record for $R=3.8 \cdot 10^5$ in
figure \ref{f4}. But in contrast to the vacillation of the right plot
figure \ref{f3} the pattern is now strongly modulated. The component with the azimuthal wavenumber m=1
plays a dominant role in this modulation. As can be seen in figure 6 the
pattern recurs nearly periodically in time in spite of the modulation.
Figure 6 corresponds to the slightly higher Rayleigh number of $4 \cdot 10^5$
where convection changes intermittently between relaxation cycles and
vacillation oscillations. 
\begin{figure}[t]
\begin{center}
\epsfig{file=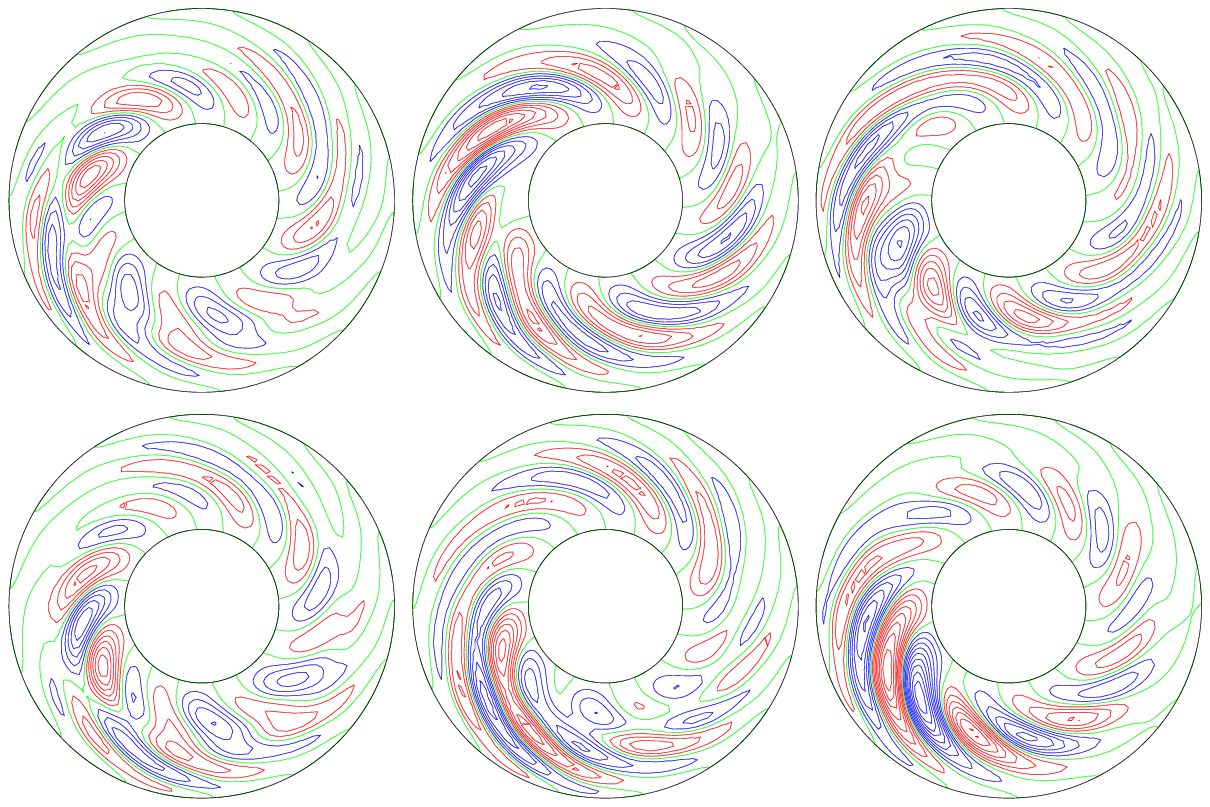,width=11cm}
\end{center}
\caption{Modulated shape vacillations of convection for $R=4 \cdot 10^5$,
$\tau=3 \cdot 10^4$, $P=0.1$. The sequence of plots equidistant in time ($\Delta t=0.00376$),
starting at the upper left and continuing clockwise, shows streamlines, $r \frac{\partial v}{\partial \phi}=const.$, in     
the equatorial plane. Since the modulation period is $t_p=0.0188$, the last plot
closely resembles the first plot except for a shift in azimuth.}
\end{figure}

As $R$ is further increased the spatio-temporal structure of convection
becomes more irregular as can be seen in the time series for $R=4.5 \cdot 10^5$ in
figure \ref{f4}. But at $R=5 \cdot 10^5$ the relaxation oscillations with only
intermittent convection become firmly established and they continue to
persist up to Rayleigh numbers of the order of $1 \cdot 10^6$. They are
basically the same phenomenon as found by Grote and Busse (2001) at $P=1$ and
even the period is the same, about $0.1$, which corresponds to the viscous
decay time of the differential rotation. The only
difference that can be noted is the small precursor hump before the burst
of convection. In figure 7 the third plot shows convection appearing in a
fairly regular pattern. Afterwards it decays again as shown in the fourth
plot before it erupts in a chaotic fashion as the differential rotation has
reached its minimum amplitude. Since this latter part of the cycle occurs
on a rather short time scale, smaller intervals between the plots have been
used for this part in figure 7.

\begin{figure}[t]

\begin{center}
\epsfig{file=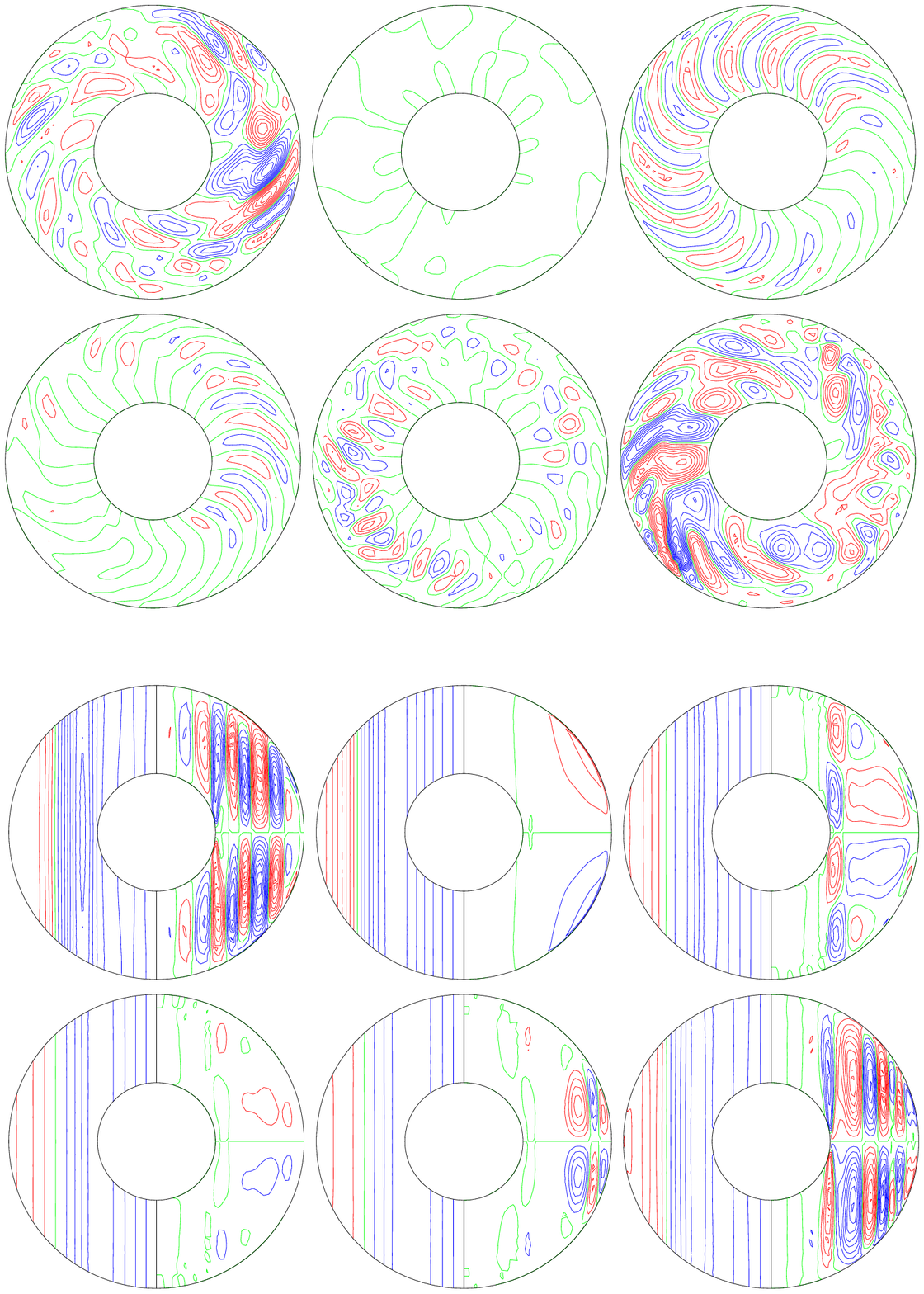,width=11cm}
\end{center}
\caption{Relaxation oscillations of convection for $R=5 \cdot 10^5$,
  $\tau=3 \cdot 10^4$, 
$P=0.1$. The upper 6 plots show a time sequence (upper row left to
right, lower row left to right) of streamlines,  $r \frac{\partial v}{\partial
  \phi}=const.$, in the equatorial plane. The 
corresponding lower 6 plots show lines of constant $u_\phi$ in the
left halves of the 
circles and the meridional circulation in the right halves. The
separation in time is 
$0.0378$ for the first four plots and $0.0126$ for the last three plots.}
\end{figure}
\begin{figure}[t]
\begin{center}
\hspace*{-0.4cm}\epsfig{file=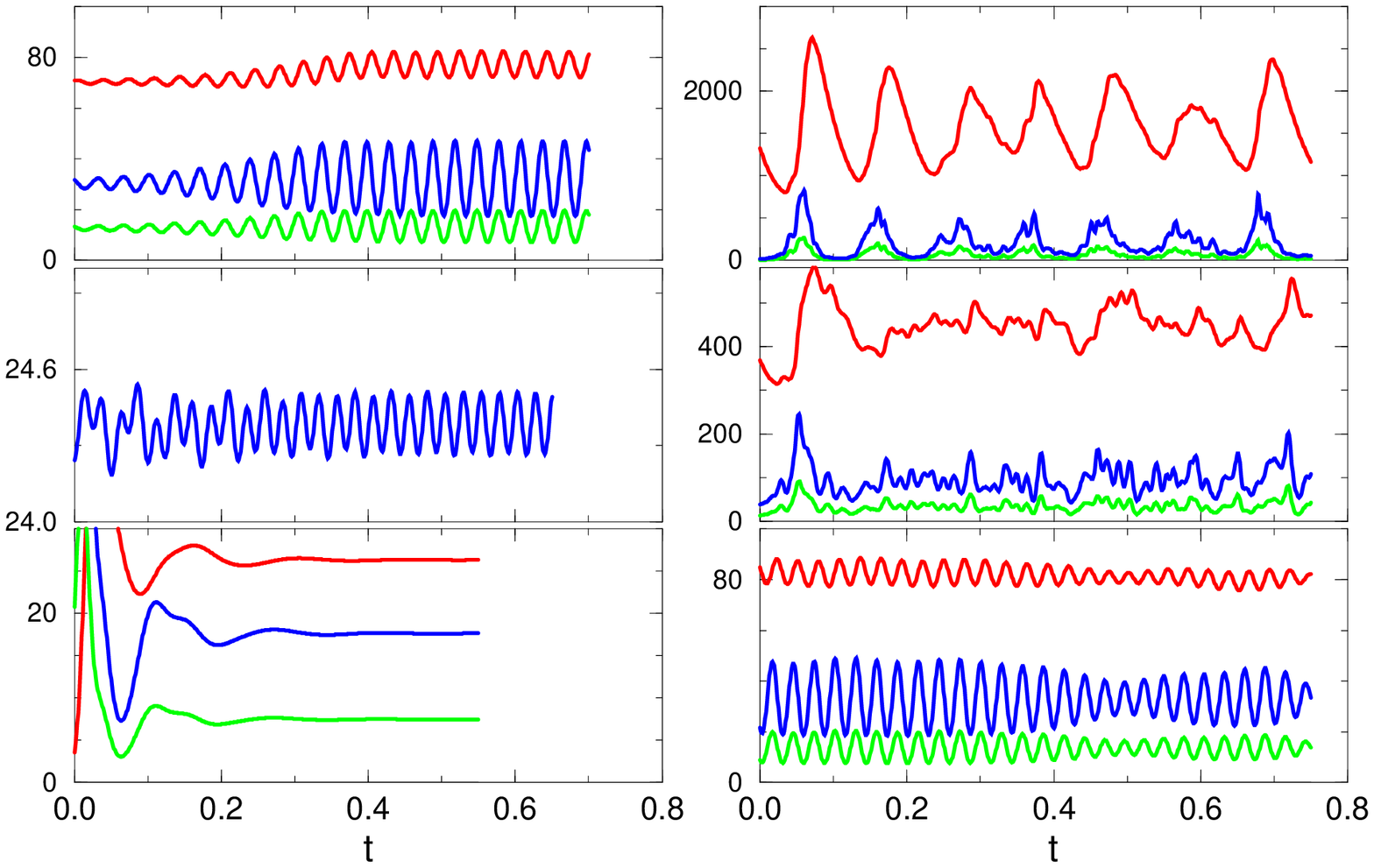,width=8cm,angle=-90}
%\hspace*{-0.4cm}\epsfig{file=fig8.eps,width=10cm,angle=-90}
\end{center}
\caption{Time series of energy densities of convection in the case
$\tau=1.5\cdot10^4$, $P=0.5$ for $R=3\cdot10^5$, $3.2\cdot10^5$,
$3.45\cdot10^5$  (from bottom to top, left side) and
$R= 3.5\cdot10^5$, $7 \cdot 10^5$, $10^6$ (from bottom to
top, right side).  Red, blue and green lines indicate
$E_t^m$,$E_t^f$, and $E_p^f$, respectively. The critical Rayleigh
number for onset is $R_c=215142$.}
%\caption{Time series of energy densities of convection in the case
%$\tau=1.5\cdot10^4$, $P=0.5$ for $R=3\cdot10^5$, $3.2\cdot10^5$,
%$3.45\cdot10^5$, $3.5\cdot10^5$ (from bottom to top, left side) and
%$R= 5\cdot10^5$, $7 \cdot 10^5$, $8\cdot10^5$, $10^6$ (from bottom to
%top, right side).  Solid, dot-dashed  and dashed lines indicate
%$E_t^m$,$E_t^f$, and $E_p^f$, respectively. The critical Rayleigh number for
%onset is $R_c=215142$.}
%\end{figure}
%\begin{figure}[t]
\begin{center}
\epsfig{file=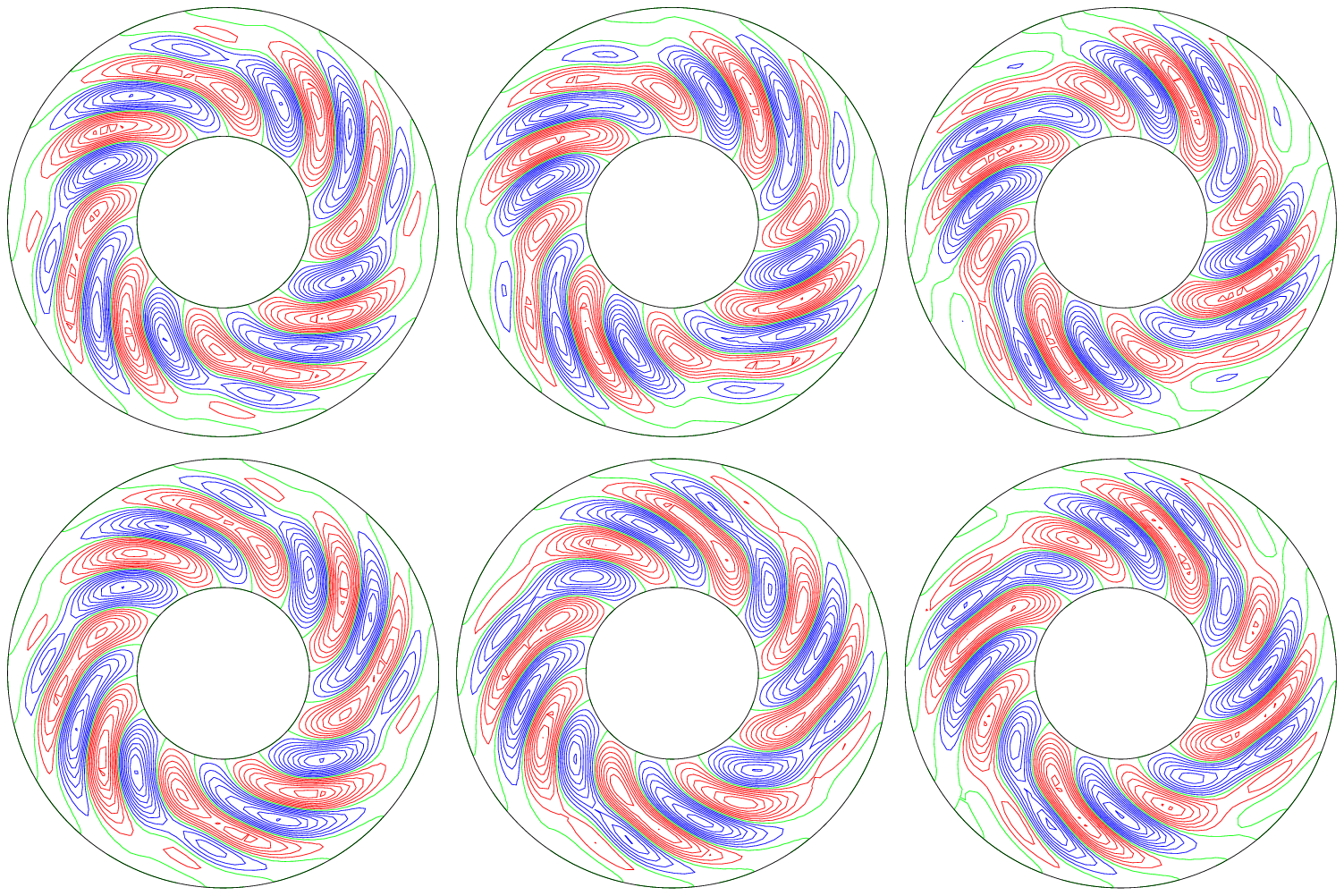,width=11cm}
\end{center}
\caption{Modulated shape vacillations of convection for $R=3.2 \cdot 10^5$,
$\tau=1.5\cdot 10^4$, $P=0.5$. The sequence of plots equidistant in time
($\Delta t=0.005$), starting at the upper left and continuing clockwise, shows
streamlines, $r \frac{\partial v}{\partial \phi}=const.$, in the equatorial plane. Since the modulation
period is about $0.025$, the last plot resembles the first plot except for a
shift in azimuth. }
\end{figure}
\begin{figure}[ht]
\begin{center}
\epsfig{file=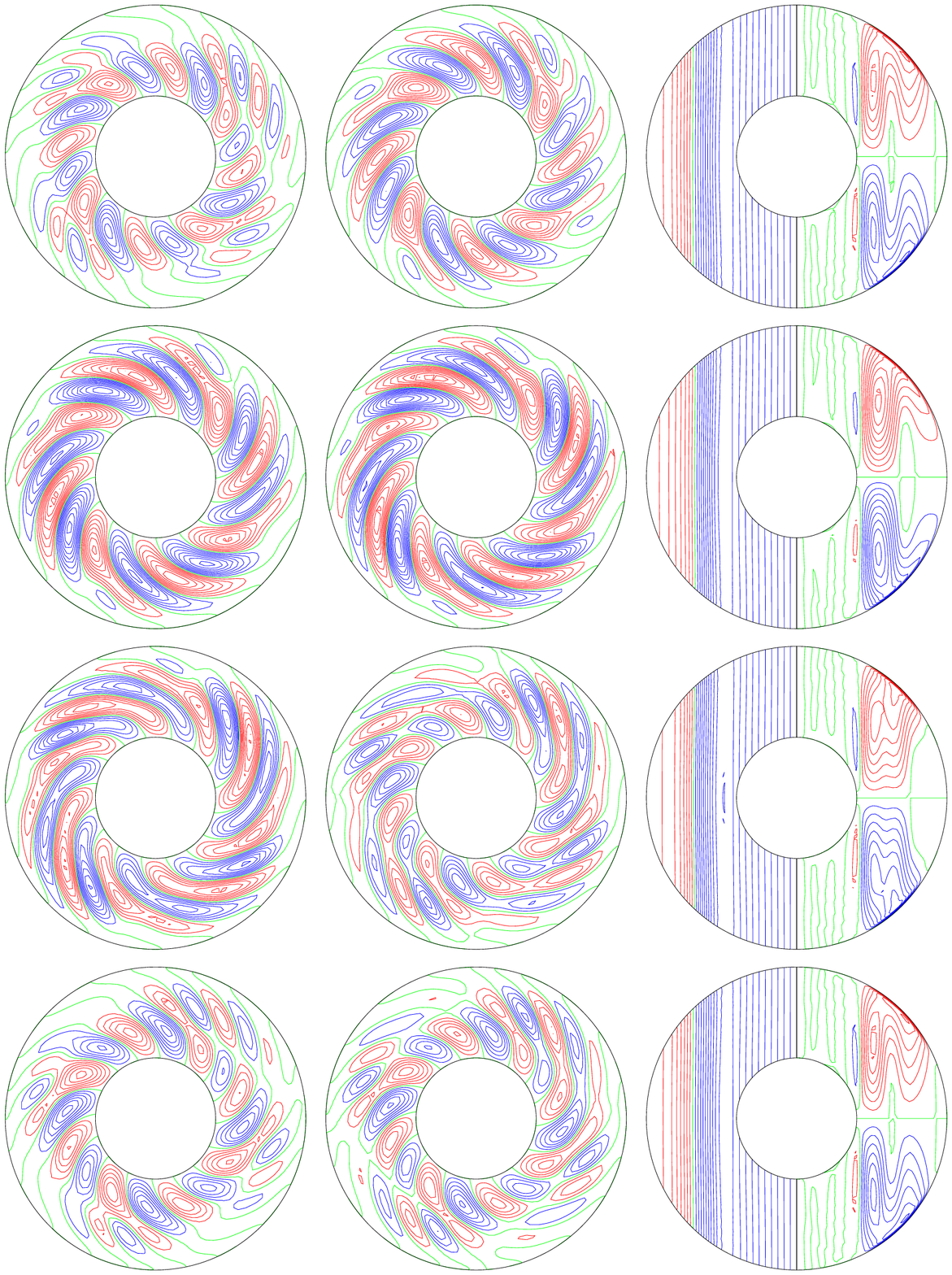,width=11cm}
\end{center}
\caption{Modulated shape vacillations of convection for $R=3.45 \cdot 10^5$,
$\tau=1.5\cdot10^4$,             $P=0.5$. The left and middle columns show
streamlines, $r \frac{\partial v}{\partial \phi}=const.$, in the equatorial plane, the right column
shows lines of constant $u_\phi$ in the left halves of the circles and
streamlines of the meridional circulation in the right halves. The left and
the right columns show plots at the times $t=(n-1) \cdot 0.01$, for $n=1,2,3,4$
(from top to bottom), the middle column shows plots at the intermediate
times $t=(n-1) \cdot 0.005$}
\end{figure}

The main difference between convection at $P=0.1$ and Prandtl numbers of the
order unity occurs at the transition between regular and irregular
patterns. A typical scenario is shown in figure 8. After convection has set
in in the form of eight drifting column pairs the usual amplitude vacillations
occur as the Rayleigh number is increased. The shape vacillations,
however, exhibit a modulation with wavenumber $m=4$ as shown in figure 9.
Only every second pair of columns gets stretched until the outer part
separates, a little earlier for the cyclonic column than for the
anticyclonic one. Then the same process is repeated for the other pairs of
columns such that the sequence shown in figure 9 exhibits only half a
period of the oscillation. In addition, of course, the column pattern
drifts in the prograde direction. Movie 2 provides a vivid impression of
this type of convection.

With increasing Rayleigh number the stretching process gets out of phase
and an $m=1$-modulation enters as shown in figure 10. The time dependence is
still periodic, but has become more complex in that some of the separated
outer parts become attached to the preceding column pair. To show this
process in more detail some extra plots have been added
to the time sequence of figure 10. While the amplitude of convection varies
considerably throughout the cycle, the differential rotation exhibits only
small oscillations as can be seen from the corresponding section of figure 8.   

By the time when $R$ has reached $3.5 \cdot 10^5$ the modulated vacillations have
become aperiodic and with increasing Rayleigh number convection becomes
more and more chaotic. Regularity reappears only in the form relaxation
oscillations as shown in the section for $R=7 \cdot 10^5$ of figure 8. Remainders
of the vacillations can still be seen in this time record. But at $R=1 \cdot 10^6$
the relaxation oscillations have become fully established with convection
occurring only in a intermittent fashion. Although obtained for a slightly
smaller value of $\tau$, movie 3 gives a good impression of the onset of
relaxation oscillations.

\begin{figure}[t]
\begin{center}
\hspace*{-1cm}\epsfig{file=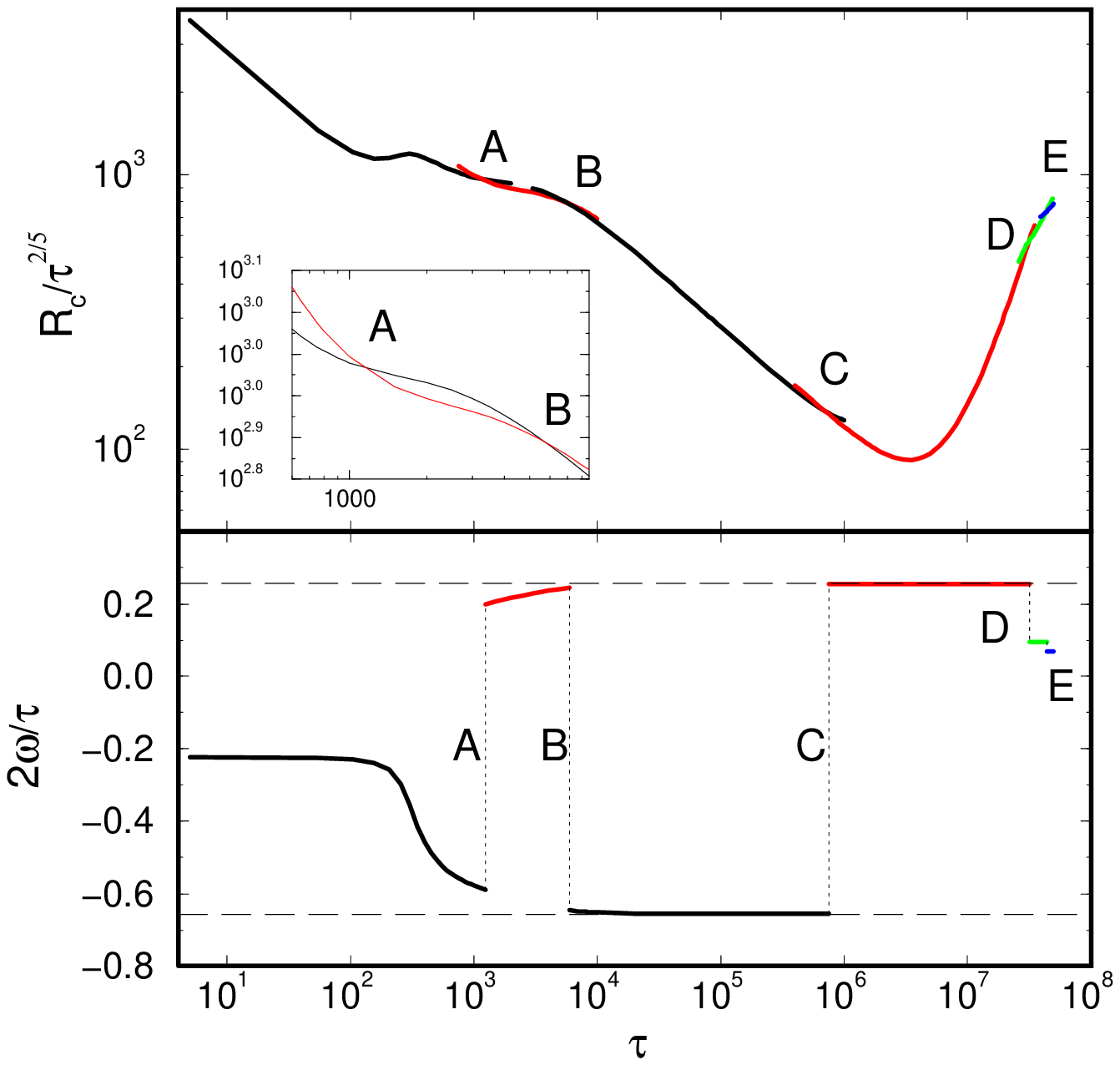,width=8cm,angle=-90}
\end{center}
\caption {The critical Rayleigh number $R_c$ (upper plot) and the
  corresponding frequency
  $\omega_c $ (lower plot) as a function of $\tau$ in the case
  $P=10^{-4}, \enspace \eta=0.3, \enspace m=8$. The values
  corresponding to the analytical expression (9)
   for $\omega_c$ are shown with dashed lines. }  
\end{figure}

\section{Convection at Small Prandtl Numbers}

Since stellar interiors as well as metallic planetary cores are
characterized by rather small Prandtl numbers much attention has been
focused on the problem of the onset of convection in rotating fluid
spheres in the case of small $P$. Zhang and Busse (1987) found the
equatorially attached mode which is quite distinct from the columnar mode
discussed in the preceding section. The new mode represents an inertial
oscillation which becomes excited when viscous dissipation is balanced by
the energy provided by thermal buoyancy. The fact that both energies can
be regarded as small perturbations has led Zhang (1994, 1995) to solve the
problem of onset of convection by an asymptotic analysis. An detailed
numerical study together with some analytical approximations can be found
in the paper of Ardes {\it et al.} (1997). 
 The results of these various efforts
have turned out to be rather complex since the preferred inertial modes
travel in the prograde as well in the retrograde directions depending on
the parameters of the problem. Moreover, the azimuthal wavenumber $m$ of
convection does not increase monotonically with the Coriolis parameter $\tau$
as is usually found for the columnar mode at values of $P$ of the order
unity or higher. In addition to the simple "single cell" inertial modes of
Zhang and Busse (1987) multicellular modes have been found in the study of 
Ardes {\it et al.} (1997) which appear to be closely related to the
multicellular modes described by the Airy function in the analysis by Yano
(1992) of the analogous problem of convection in the rotating cylindrical
annulus. The numerical solutions for the problem obtained by Pino  {\it et al.}
(2000) also indicate the onset of multicellular convection in parts of the
parameter space.
\begin{figure}[t] 
\begin{center}
\epsfig{file=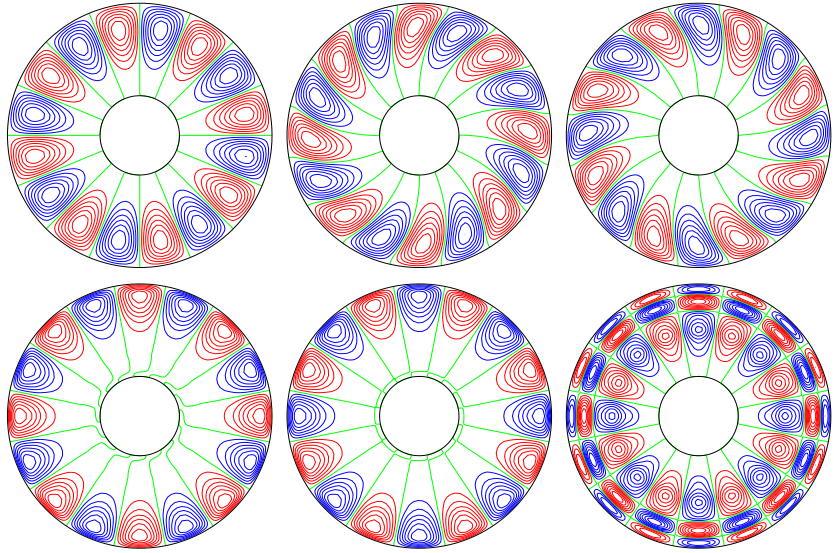,width=11cm}
\end{center}
\caption {Lines of constant $r \frac{\partial v}{\partial \phi}$ in the
  equatorial plane illustrating the sequence of transitions for the
  same parameters as Fig. 11 and  values of $\tau= 5$, $950$, $1500$, $6 \cdot
  10^5$, $8 \cdot 10^5$, $3.5\cdot 10^7 $ (left to right, first upper row then lower row). }
\end{figure}

\begin{figure}[htb] 
\begin{center}
\epsfig{file=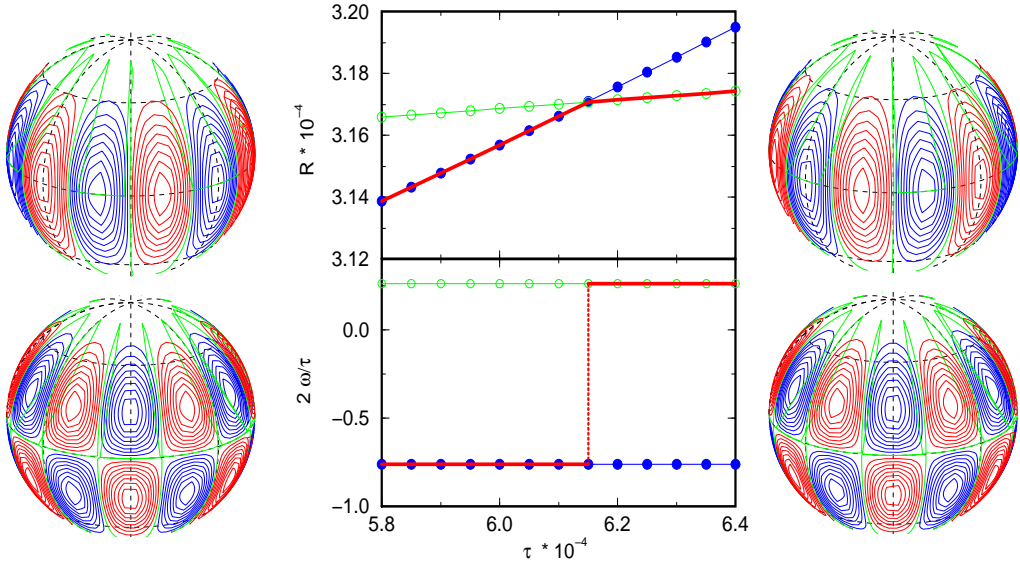,height=13.5cm,angle=-90}
\end{center}
\caption {Graphs in the center show the critical Rayleigh numbers
  $R_c$ of the competing prograde (green line,
  empty circles) and retrograde mode (blue line, filled circles)
  as well as the actual critical value (thick red line) and the
  corresponding frequencies as a function 
  of $\tau$ in the case $P=0.001$, $ \eta=0.2$,$ m=6$.
  Contours of constant radial velocity $u_r$ (lower plots) and the streamlines,
$w=const.$ on the spherical surface $r=0.9$ (upper plots) for
  $\tau=58000$ and $64000$ are shown on the left and right sides, respectively.} 
\end{figure}
\begin{figure}[htb]
\begin{center}
\epsfig{file=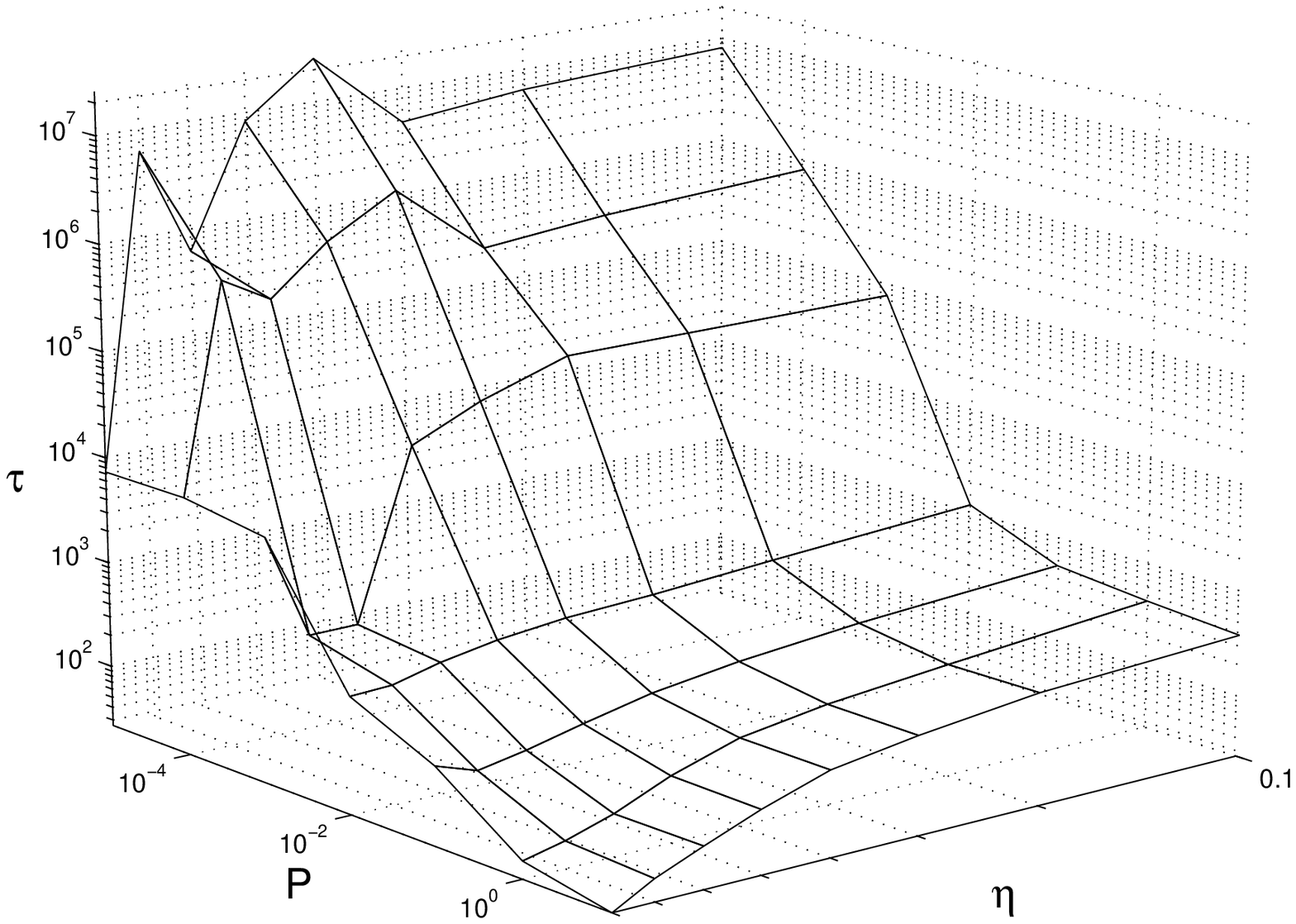,width=12cm}
\end{center}
\caption {The boundary separating the retrograde (below the surface) and
  prograde mode (above the surface) for a fixed value of the wavenumber $m=6$.}
\end{figure}      

Here we shall present only some examples of the variety of patterns
encountered in a rotating sphere of a low Prandtl number fluid. In figure
11 the multiple changes from retrograde to prograde modes as a function of
$\tau$ are indicated and the corresponding changes in the patterns are shown
in figure 12. To facilitate comparisons the wavenumber $m=8$ has been kept
fixed. But similar changes in the structure of convection are found when
competing values of $m$ are admitted. The retrograde and prograde modes
differ little in their form. The opposite phase between toroidal and
poloidal components of motion is the most characteristic difference as
indicated in figure 13. The sense of outward spiralling is also opposite
for prograde and retrograde modes as is evident from figure 12. But in the
limit of vanishing Prandtl number the sense of spiralling disappears
because the phase of the inertial modes does not vary with distance from
the axis.

%The drift freqencies of convection are essentially identical with those of
%the corresponding inertial oscillations,
%\begin{equation}
%\frac{\omega}{\tau} = \frac{1}{m+2} ( 1 \pm [ (1 + m(m+2) (2m+3)^{-1} ]^{\frac{1}{2}}).
%\end{equation}
%These expressions for prograde (negative sign) and retrograde (positive
%sign) modes are the same as those used in the paper of Zhang (1994).

Whenever $\tau$ is sufficiently large the frequency omega is closely
approximated by the frequency of the corresponding inertial modes which is
given by the analytical expression (Zhang, 1994; Ardes {\it et al.}, 1997),
\begin{equation}
\omega = \frac{\tau}{m+2} ( 1 \pm [ (1 + m(m+2) (2m+3)^{-1} ]^{\frac{1}{2}}).
\end{equation}

\begin{figure}[t]
\begin{center}
\epsfig{file=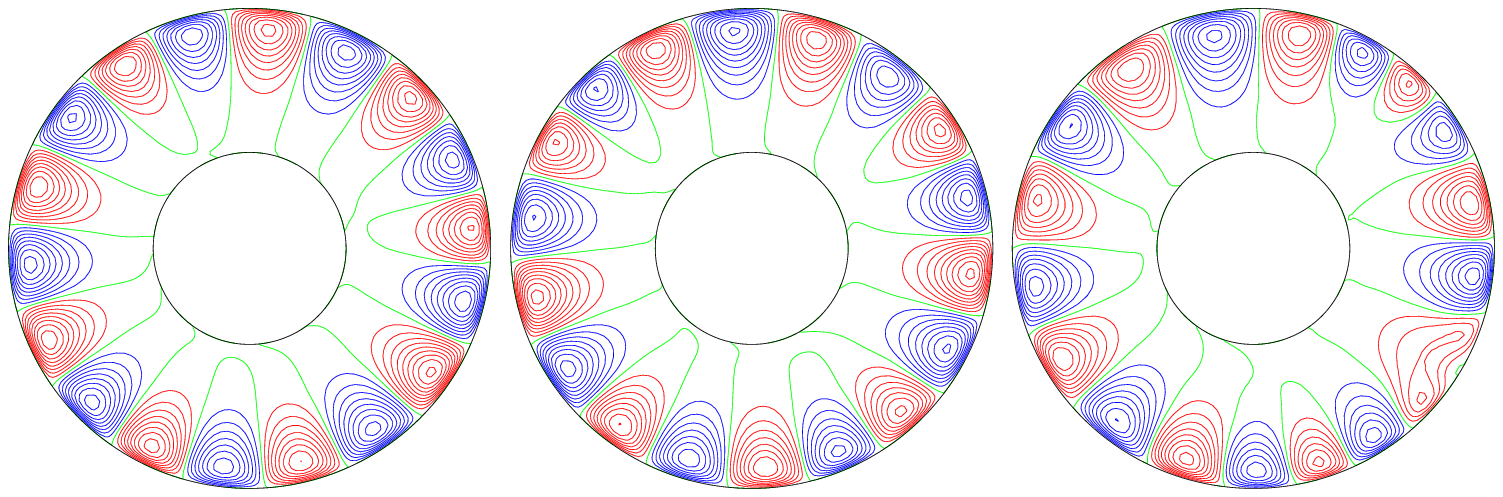,width=11cm}
\end{center}
\caption{Streamlines $r \frac{\partial v}{\partial \phi}=const.$ in
the equatorial plane for the case $P=0.025$, $\tau=10^5$ with $R=3.2
\cdot 10^5$, $3.4 \cdot 10^5$, $3.8 \cdot 10^5$ (from left to right).}
\end{figure}
\begin{figure}[htb]
\begin{center}
\epsfig{file=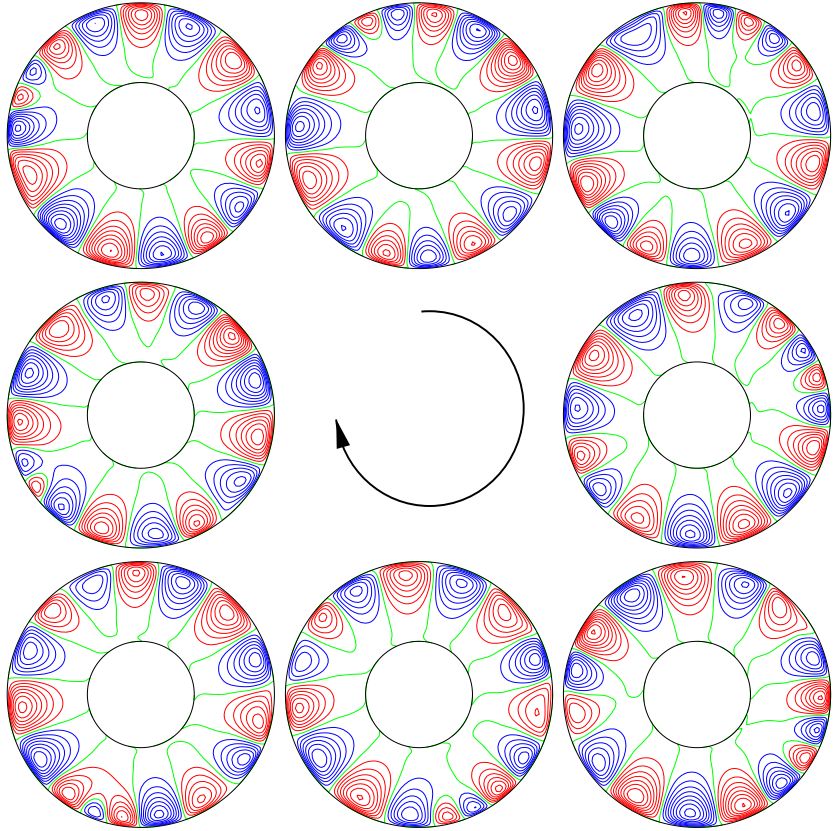,width=11cm}
\end{center}
\caption{Streamlines $r \frac{\partial v}{\partial \phi}=const.$ in
the equatorial plane for the case  $P=0.025$, $\tau=10^5$, $R=4 \cdot
10^5$. Time increases by $4 \cdot 10^{-3}$ from one plot to the next
starting at the left upper corner. The eight plot is similar to the
first one except for a shift in azimuth.}
\end{figure}

\begin{figure}[htb]
\begin{center}
\hspace*{0cm}\epsfig{file=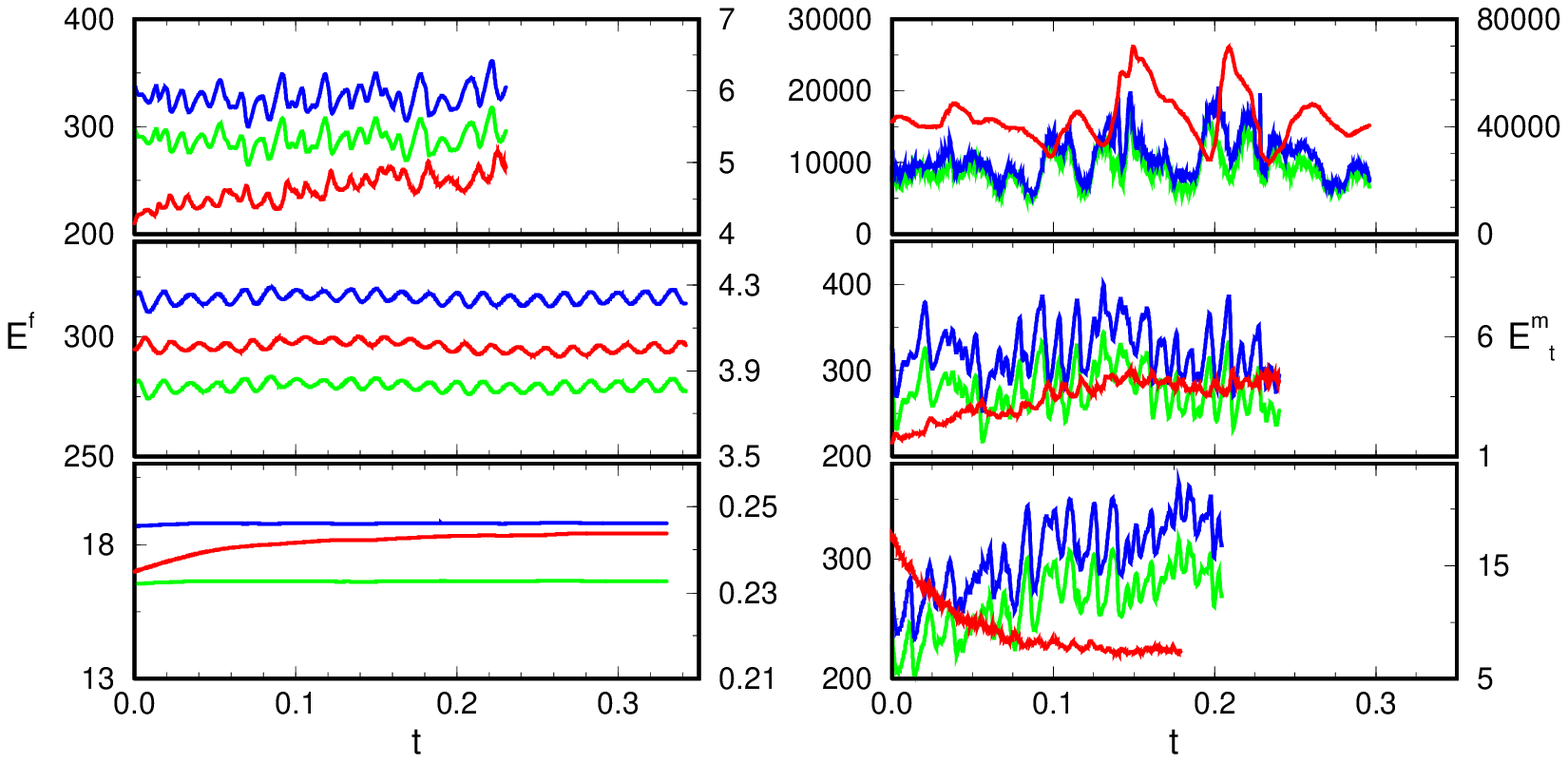,width=15cm}
\end{center}
\caption{Time series of energy densities of convection in the case
$P=0.025$, $\tau=10^5$,
for $R=3.1 \cdot 10^5$, $3.2 \cdot 10^5$, $3.4 \cdot 10^5$, (from
bottom to top, left side) and $3.8 \cdot 10^5$, $4 \cdot 10^5$, $8
\cdot  10^5$  (from bottom to
top, right side).  The red, blue and green
lines correspond to $E^m_t$,  $E^f_t$, $E^f_p$ , respectively. $E^m_t$
is measured on the left ordinate.} 
\end{figure}
The negative (positive) sign applies for modes drifting in the prograde
(retrograde) direction. Explicit values of expression (9) are given in
figures 11 and 13 for comparisons with the numerically determined values.

For an analysis of nonlinear properties of equatorially attached convection
we focus on the case $P=0.025$ with $\tau=10^5$. The critical Rayleigh number
for this case is $R_c=28300$ corresponding to $m=10$. As R is increased beyond
the critical value other values of $m$ from 7 to 12 can be realized, but $m=10$
and lower values are usually preferred. An asymptotic perfectly periodic
solution with $m=10$ or $m=9$ can be found only for Rayleigh numbers close to
the critical value when computations are started from arbitrary initial
conditions. On the other hand, perfect periodic patterns appear to be
stable with respect to small disturbances over a more extended regime of
supercritical Rayleigh numbers. Distinct transitions like the transition to
amplitude vacillations and to structure vacillations do not seem to exist
for equatorially attached convection. Instead modulated patterns are
typically already observed when $R$ exceeds the critical value by 10\% as can
be seen in the plots of figure 15.
These modulations are basically caused by the interaction of two modes with
neighboring values of the azimuthal wavenumber $m$ as indicated in figure 16.
Here the region of small amplitude drifts in the retrograde direction with
the difference of the drift rates for $m=7$ and $m=8$.
The time series of energy densities shown in figure 17 indicate that
more   than two modes usually contribute to the dynamics of the pattern since the
time dependence is not periodic as one would expect if only two modes
interact. The computations of the time series require a high spatial
resolution together with a small time step. The time spans indicated in
figure 17 are sufficient for reaching a statistically steady state of the
fluctuating components of motion since these equilibrate on the fast
thermal time scale of the order $P^{-1}$. Only close to $R_c$ the adjustment
process takes longer as can be seen in the case $R=3.1 \cdot 10^5$ where a
$m=10$ - pattern approaches its equilibrium state. The pattern corresponding to
the other cases of figure 17 are the ones shown in figures 15, 16 and 18. The
differential rotation represented by $E_t^m$ relaxes on the viscous time
scale and thus takes a long time to reach its asymptotic regime in the
examples shown in figure 17. But the differential rotation is quite weak such that it has a
negligible effect on the other components of  motion except in the
case of the highest Rayleigh number of figure 17. Even smaller is
the axisymmetric part of the poloidal component of  motion which is
not shown in the plots of figure 17.  At higher Rayleigh numbers the
equatorially attached eddies spread  farther into the interior and in
some cases become detached from the equator as can be seen in the
plots of figure 18. In this way the convection eddies contribute to
the heat transport from the inner boundary. But at the same time they
acquire the properties of the convection columns which are
characteristic for convection at higher Prandtl numbers. Accordingly
the differential rotation is steeply increased at $R=10^6$ and a tendency
towards relaxation oscillation can be noticed in the upper right time series of
figure 17.

\section{Concluding Remarks}

The results on linear and nonlinear properties of low Prandtl number
convection in rotating spherical shells are of special interest for
applications to problems of convection in planetary and stellar interiors.
Since planetary cores often consist of liquid metal or of metallic hydrogen
as in the case of Jupiter and Saturn, their Prandtl numbers are rather low.
Moreover, because of high temperatures radiative heat transport is no
longer negligible which tends to lower the Prandtl number even more. The
latter effect dominates in stellar interiors.
On the other hand, turbulence owing to small scale motions which remain
unresolved in numerical simulations will tend to equalize all effective
diffusivities. But this tendency is not likely to erase entirely the
differences in the subgrid scale transports of heat and momentum.
\begin{figure}[ht]
\begin{center}
\hspace*{0cm}\epsfig{file=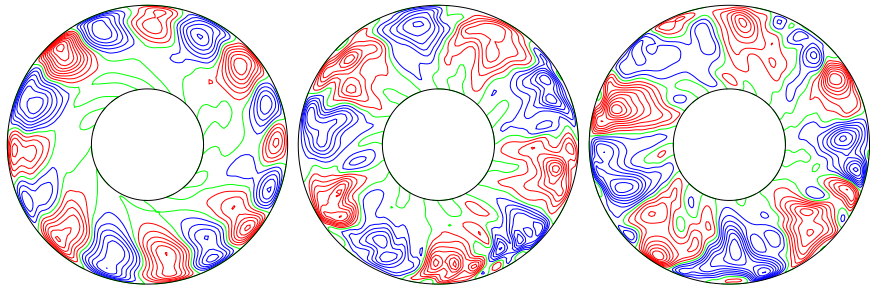,width=11.5cm}
\end{center}
\caption{Streamlines $r \frac{\partial v}{\partial \phi}=const.$ in
the equatorial plane for the case $P=0.025$, $\tau=10^5$ with $R=6
\cdot 10^5$, $8 \cdot 10^5$, $10^6$ (from left to right).}
\end{figure}

    Of particular interest is the effect of the Prandtl number on the
dynamo action of convection. While the strong differential rotation
associated with columnar convection tends to promote the generation of
magnetic fields (see, for example, Grote and Busse, 2001), its near absence
in the case of equatorial attached convection may increase the Rayleigh
number required for dynamo action at a given value of the magnetic Prandtl
number. As the Rayleigh number reaches high values more column like convection
flows tend to appear near the inner boundary of the shell and are expected
to promote dynamo action. Numerical studies of this problem are in
progress and will be reported in the near future.
 
 The regular and modulated patterns of convection reported in this
 paper as well as the coherent structures found in the chaotic regimes
 of convection can in principle be observed in experiments through the
 use of techniques mentioned in the Introduction. Low Prandtl number
 fluids such as liquid metals pose severe experimental challenges
 because convection flows 
 can not  easily be visualized. On the other hand the temperature
 profiles of convection patterns can be measured through time
 records of thermistor probes implanted in the spherical boundary since
 the pattern is drifting by. See, for example, measurements of
 convection in mercury in the case of the cylindrical annulus by Azouni
 {\it et al.} (1986) and of convection in a spherical shell by Cordero
 and Busse (1992). For the purpose of an eventual comparison with
 experimental observation the computations reported in this paper
 should be repeated for the case of no-slip boundaries with $R_e \ne
 0$ and $R_i = 0$. The computational effort will be more demanding but
 no significant qualitative changes are expected in
 that case.

\section*{References}
\begin{itemize}
\item[]
Ardes, M., Busse, F.H., and Wicht, J., Thermal Convection in Rotating
Spherical Shells, {\it Phys. Earth Plan. Int.} {\bf 99}, 55-67, 1997
\item[]
Azouni, A., Bolton, E.W., and Busse, F.H., Experimental study of
convection in a rotating cylindrical annulus, {\it Geophys. Astrophys.
  Fluid Dyn.}, {\bf 34}, 301-317, 1986
\item[]
Busse, F.H.: Thermal instabilities in rapidly rotating systems. {\it
  J. Fluid Mech.}, 
{\bf44}, 441-460, 1970
%\item[]
%Busse, F.H., Differential Rotation in Stellar Convection Zones, {\it Astrophys.
%J.}, {\bf 159}, 629-639, 1970b
\item[]
Busse, F.H., A simple model of convection in the Jovian atmosphere,
{\it Icarus} {\bf 20}, 255-260, 1976
\item[]
Busse, F.H., Convection driven zonal flows and vortices in the major
planets, {\it CHAOS}, {\bf 4}, 123-134, 1994
\item[]
Busse, F.H., Convection  flows in rapidly rotating spheres and their
dynamo action, {\it Phys. Fluids}, {\bf 14}, 1301-1314, 2002
\item[]
Busse, F.H., Carrigan, C.R., Laboratory simulation of thermal
convection in rotating planets and stars, {\it SCIENCE}, {\bf 191},
81-83, 1976
\item[]
Busse, F.H., Grote, E., Tilgner, A., On convection driven dynamos in
rotating spherical shells. {\it Studia geoph. et geod.}, {\bf  42} , 211-223, 1998
\item[]
Christensen, U.R. {\it et al.}, A numerical dynamo benchmark. {\it Phys. Earth Plan. Int.},
{\bf 128}, 25-34, 2001
\item[]
Cordero, S., and Busse, F.H., Experiments on convection in rotating
hemispherical shells: Transition to a quasi-periodic state, {\it
  Geophys. Res. Letts.}, {\bf  19}, 733-736, 1992
\item[]
Grote, E., and Busse, F.H., Dynamics of Convection and Dynamos in Rotating
Spherical Fluid Shells, {\it Fluid Dyn. Res.}, {\bf 28}, 349-368, 2001
\item[]
Hart, J.E., Glatzmaier, G.A.,Toomre, J., Space-laboratory and
numerical simulations of thermal convection in a rotating
hemi-spherical shell with radial gravity, {\it J. Fluid Mech.}, {\bf
173}, 519-544, 1986 
\item[]
Jones, C.A., Soward, A.M., and Mussa, A.I., The onset of thermal
convection in a rapidly rotating sphere,  {\it J. Fluid Mech.}, {\bf
405}, 157-179, 2000
\item[]
Pino, D., Mercader, I., and  Net, M., Thermal
and inertial mode of convection in a rapidly rotating annulus, {\it
  Phys. Rev. E}, {\bf 61}, 1507-1517, 2000 
%Pino, D., Net, M., S'anchez, J., and Mercader, I., Thermal
%Rossby waves in a rotating annulus. Their stability, {\it
%Phys. Rev. E}, {\bf 63}, 056312, 2001. 
\item[]
Rosenzweig, R.E., Browaeys, J., Bacri, J.-C., Zebib, A., and Perzynski, R., 
Laboratory study of Spherical Convection in Simulated Central Gravity,  
{\it Phys. Rev. Letts..}, {\bf 83}, 4904-4907, 1997
\item[]
Tilgner, A., and Busse, F.H., Finite amplitude convection in rotating 
spherical fluid shells, {\it J. Fluid Mech.}, {\bf 332}, 359-376, 1997
\item[]
Yano, J.-I., Asymptotic theory of thermal convection in rapidly rotating
systems, {\it J.Fluid Mech.},{\bf 243}, 103-131, 1992  
\item[]
Zhang, K., Vacillating convection in a rotating spherical fluid shell
at infinite Prandtl number,  {\it J. Fluid Mech.}, {\bf 228}, 607-628, 1991
\item[]
Zhang, K., Spiraling columnar convection in rapidly rotating spherical
fluid shells, {\it J. Fluid Mech.}, {\bf 236}, 535-556, 1992
\item[]
Zhang, K., On coupling between the Poincar\'e equation and the heat
equation, {\it J. Fluid Mech.}, {\bf 268}, 211-229, 1994
\item[]
Zhang, K., On coupling between the Poincar\'e equation and the heat
equation: no-slip boundary condition, {\it J. Fluid Mech.}, {\bf 284},
239-256, 1995 
\item[]
Zhang, K., and Busse, F.H., On the onset of convection in rotating
spherical shells, {\it Geophys. Astrophys. Fluid Dyn.}, {\bf 39},
119-147, 1987
\item[]
Zhang, K., and Busse, F.H., Some recent developments in the theory of
convection in rotating systems, in ``Nonlinear Instability, Chaos
and Turbulence'', Vol 1, L. Debnath and D.N. Riahi, eds., {\it
Computational Mechanics Publications, WIT Press}, 17-69, 1998 
\end{itemize}
\end{document}